# Fluctuation conductivity due to the preformed local pairs

T. Domański[*]

*Institute of Physics, M. Curie Skłodowska University, 20-031 Lublin, Poland*

M. Barańska

*Institute of Physics, Polish Academy of Sciences, 02-668 Warsaw, Poland*

A.L. Solovjov

*B. Verkin Institute for Low Temperature Physics and Engineering of
the National Academy of Sciences of Ukraine, 61103 Kharkov, Ukraine*

(Dated: April 4, 2018)

We investigate the properties of a system where the itinerant electrons coexist and interact with the preformed local pairs. Using the nonperturbative continuous unitary transformation technique we show that Andreev-type scattering between these charge carriers gives rise to the enhanced diamagnetic response and is accompanied by appearance of the Drude peak inside the pseudogap regime $\omega \leq 2\Delta_{pg}$. Both thes effects are caused above the transition temperature $T_c$ by the short-range superconducting correlations. In fact, the residual diamagnetism has been detected by the torque magnetometry in the lanthanum and bismuth cuprate superconductors at temperatures up to $\sim 1.5 T_c$. In this work we show how the superconducting correlations can be observed in the a.c. and d.c. conductivity.

## I. INTRODUCTION

One of important aspects concerning the role of electron correlations in the cuprate oxides refers to the pseudogap phase, (existing above the transition temperature $T_c$) and its relationship with the true superconducting state [1]. Origin of the entire pseudogap state is still a matter of controversy [2], but a number of experimental data [3–12] clearly indicate that the superconducting correlations emerge gradually upon approaching $T_c$ from above. Precursor signatures are seen e.g. in a weak diamagnetic response above the superconducting dome (reported by the torque magnetometry [4]) or the short-scale superconducting correlations (detected by the early ultrafast spectroscopy [5] and the recent transient effects [13, 14]). Physically these effects are driven by the preformed pairs which are correlated above $T_c$ only on some finite spatial and/or temporal scales.

Consequences of the short-range correlated preformed pairs can be also probed by the finite-frequency optical conductivity. Rich experimental data on the electrodynamic properties [15, 16] have been so far discussed in terms of the extended Drude model, determining the frequency dependent relaxation time $\tau(\omega)$. Interpretation of the precursor effects within such framework is rather complicated because, on one hand the depleted single-particle spectrum suppresses the subgap optical weight, and on the other hand appearance of the pair correlations gives rise to the zero-frequency Drude peak [17], signalling a fragile superfluid stiffness. Similar fluctuation effects have been also reported for the thin samples of the strongly disordered $s$-wave superconductors [18]. These physical processes have been studied within the diagrammatic approximation for the current-current response function, using the dressed single particle propagators [19–21], imposing the selfconsistent conserving scheme [22] or inventing other sophisticated methods for the vertex corrections [23, 24].

In this paper we address qualitative changes of the conductivity driven by the preformed pairs, going beyond the usual perturbative framework. For this purpose we adopt phenomenological scenario describing the local (preformed) pairs coexisting and interacting with single (unpaired) electrons. We treat on equal footing the boson and fermion degrees of freedom, by means of the continuous unitary transformation [25, 26] that is reminiscent of the numerical renormalization group techniques [27]. Such nonperturbative scheme has been used by us [28] to determine the response function beyond the BCS approximation [29]. Here we focus on its physical implications for the real part of the frequency-dependent conductance due to the preformed local pairs. In particular, we show that the Drude-like feature appears in the subgap (infrared) regime and it acquires more and more spectral weight upon approaching $T_c$ from above. We confront this prediction with the experimental data obtained for Bi2223 cuprates.

## II. PREFORMED PAIRS SCENARIO

Effects of the preformed pairs (of whatever origin) can be studied using the boson-fermion Hamiltonian

$$\hat{H} = \sum_{\mathbf{k},\sigma} \xi_{\mathbf{k}} \hat{c}^{\dagger}_{\mathbf{k}\sigma} \hat{c}_{\mathbf{k}\sigma} + \sum_{\mathbf{q}} E_{\mathbf{q}} \hat{b}^{\dagger}_{\mathbf{q}} \hat{b}_{\mathbf{q}} \qquad (1)$$
$$+ \frac{1}{\sqrt{N}} \sum_{\mathbf{k},\mathbf{p}} g_{\mathbf{k},\mathbf{p}} \left( \hat{b}^{\dagger}_{\mathbf{k}+\mathbf{p}} \hat{c}_{\mathbf{k}\downarrow} \hat{c}_{\mathbf{p}\uparrow} + \hat{b}_{\mathbf{k}+\mathbf{p}} \hat{c}^{\dagger}_{\mathbf{k}\uparrow} \hat{c}^{\dagger}_{\mathbf{p}\downarrow} \right).$$

[*]Electronic address: doman@kft.umcs.lublin.pl

This model describes the itinerant electrons (fermion operators $\hat{c}^{(\dagger)}_{\mathbf{k}\sigma}$) coexisting with the tightly bound pairs (boson operators $\hat{b}^{(\dagger)}_{\mathbf{q}}$), where $\xi_{\mathbf{k}}$ measures the energy with respect to chemical potential $\mu$ and $E_{\mathbf{q}}$ is the energy of preformed pairs measured with respect to $2\mu$. For treating the Bose-Einstein (BE) condensed pairs (i.e. $\mathbf{q}=\mathbf{0}$ mode) one can simplify (1) to the standard BCS Hamiltonian with $\Delta_{\mathbf{k}} = -\frac{\hat{b}_{\mathbf{q}=\mathbf{0}}}{\sqrt{N}}\, g_{\mathbf{k},-\mathbf{k}}$. It is the purpose of our study here to address the role of finite momentum pairs $\hat{b}_{\mathbf{q}\neq\mathbf{0}}$.

Specific argumentation in favor for the boson-fermion scenario (1) has been discussed by various groups [30–37]. This Hamiltonian can be derived from the plaquettized Hubbard model using the contractor method [33]. Such model has been shown [34] to capture the Anderson's idea of the resonating valence bond picture. The Hamiltonian (1) has been also deduced on phenomenological grounds [35–37] as realistic prototype for the correlated electrons (holes) in $CuO_2$ planes. It also accounts for the resonant Feshbach interaction operating in the ultracold fermion atoms such as $^6$Li or $^{40}$K [38–40].

## III. SINGLE-PARTICLE VS COLLECTIVE FEATURES

For studying influence of the preformed pairs on the single-particle electron spectrum (and *vice versa*) we construct the unitary transformation $\hat{U}(l)$, diagonalizing the Hamiltonian $H(l) = U^{\dagger}(l)HU(l)$ in a continuous manner. The transformed Hamiltonian $H(l)$ evolves with respect to a formal parameter $l$ via the flow equation [25, 26]

$$\frac{d\hat{H}(l)}{dl} = [\hat{\eta}(l), \hat{H}(l)] \qquad (2)$$

with the generating operator $\hat{\eta}(l) \equiv \frac{d\hat{\mathcal{U}}(l)}{dl}\hat{\mathcal{U}}^{-1}(l)$. Hamiltonians $\hat{H}(l) = \hat{H}_0(l) + \hat{H}_1(l)$ (where $\hat{H}_0(l)$ describes the diagonal part and $\hat{H}_1(l)$ is the off-diagonal term) can be asymptotically diagonalized

$$\lim_{l\to\infty} \hat{H}_1(l) = 0 \qquad (3)$$

applying the following generating operator [25]

$$\hat{\eta}(l) = \left[\hat{H}_0(l), \hat{H}_1(l)\right]. \qquad (4)$$

During the unitary transformation all the model parameters are continuously renormalized to their asymptotic (fixed point) values [26]. Adopting this algorithm (4) we have constructed [41, 42] the continuous unitary transformation for the model (1), choosing $\hat{H}_0(l) = \sum_{\mathbf{k},\sigma}\xi_{\mathbf{k}}(l)\hat{c}^{\dagger}_{\mathbf{k}\sigma}\hat{c}_{\mathbf{k}\sigma} + \sum_{\mathbf{q}} E_{\mathbf{q}}(l)\hat{b}^{\dagger}_{\mathbf{q}}\hat{b}_{\mathbf{q}}$ and $\hat{H}_1(l) = \hat{H}(l) - \hat{H}_0(l)$. The generating operator (4) is then given by

$$\hat{\eta}(l) = \frac{1}{\sqrt{N}}\sum_{\mathbf{k},\mathbf{p}}\left(\alpha_{\mathbf{k},\mathbf{p}}(l)\hat{b}_{\mathbf{k}+\mathbf{p}}\hat{c}^{\dagger}_{\mathbf{k}\uparrow}\hat{c}^{\dagger}_{\mathbf{p}\downarrow} - \text{h.c.}\right), \qquad (5)$$

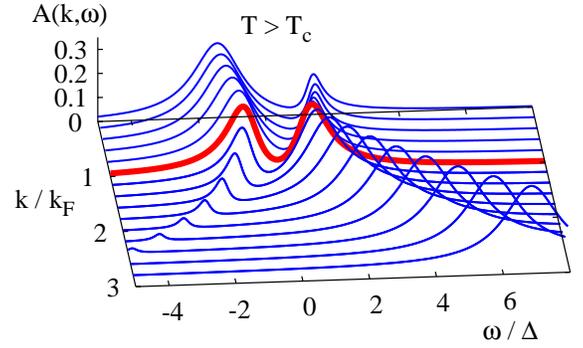

FIG. 1: (Color online) Schematic view of the gaped fermion spectrum with the Bogolubov-type quasiparticle branches surviving above $T_c$. Results are obtained for the boson-fermion model (1) using the procedure discussed in Ref. [43].

where $\alpha_{\mathbf{k},\mathbf{p}}(l) = g_{\mathbf{k},\mathbf{p}}(l)[\xi_{\mathbf{k}}(l) + \xi_{\mathbf{p}}(l) - E_{\mathbf{k}+\mathbf{p}}(l)]$. Substituting (5) to the flow equation (1) one obtains [41]

$$\frac{d}{dl}\ln g_{\mathbf{k},\mathbf{p}}(l) = -\left[\xi_{\mathbf{k}}(l) + \xi_{\mathbf{p}}(l) - E_{\mathbf{k}+\mathbf{p}}(l)\right]^2. \qquad (6)$$

This equation (6) implies an exponential decay of the boson-fermion coupling $g_{\mathbf{k},\mathbf{p}}(l)$ in the limit $l \to \infty$. Simultaneously, the fermion and boson energies are renormalized according to the flow equations [41]

$$\frac{d}{dl}\xi_{\mathbf{k}}(l) = \frac{2}{N}\sum_{\mathbf{q}}\alpha_{\mathbf{k},\mathbf{q}-\mathbf{k}}(l)\, g_{\mathbf{k},\mathbf{q}-\mathbf{k}}(l)\, n^B_{\mathbf{q}} \qquad (7)$$

$$\frac{d}{dl}E_{\mathbf{q}}(l) = -\frac{2}{N}\sum_{\mathbf{k}}\alpha_{\mathbf{k},\mathbf{k}-\mathbf{q}}(l)\, g_{\mathbf{k}-\mathbf{q},\mathbf{k}}(l)$$
$$\times \left[1 - n^F_{\mathbf{k}-\mathbf{q},\uparrow} - n^F_{\mathbf{k},\downarrow}\right] \qquad (8)$$

where $n^F_{\mathbf{k},\sigma}$ ($n^B_{\mathbf{q}}$) denotes the fermion (boson) occupancy. We have selfconsistently solved the equations (6-8) for the fixed charge concentration $n_{tot} = \sum_{\mathbf{k},\sigma}n^F_{\mathbf{k},\sigma} + 2\sum_{\mathbf{q}}n^B_{\mathbf{q}}$. The (asymptotic) dispersions $\tilde{\xi}_{\mathbf{k}} \equiv \lim_{l\to\infty}\xi_{\mathbf{k}}(l)$ and $\tilde{E}_{\mathbf{q}} \equiv \lim_{l\to\infty}E_{\mathbf{q}}(l)$ revealed that [41, 42]:

a) the fermionic spectrum is gaped around $\mu$ with the Bogolubov-type quasiparticle branches existing below and above $T_c$ (see Fig. 1),

b) the low-energy bosonic spectrum is characterized above $T_c$ by the parabolic function $(\hbar\mathbf{q})^2/2m^B$ with temperature-dependent effective mass $m^B$ (Fig. 2) which evolves at temperatures $T < T_c$ into the sound-wave Goldstone dispersion $v_s|\mathbf{q}|$.

We would like to emphasize that the Bogolubov quasiparticle branches surviving above $T_c$ have been later on confirmed experimentally by the angle resolved photoemission spectroscopy for the bismuth [9] and lanthanum compounds [10]. Similar effect has been also reported by the **k**-resolved radiofrequency spectroscopy for the ultracold potassium atoms [44]. This typical superconducting feature has been observed in the normal state even in absence of the long-range pair coherence.

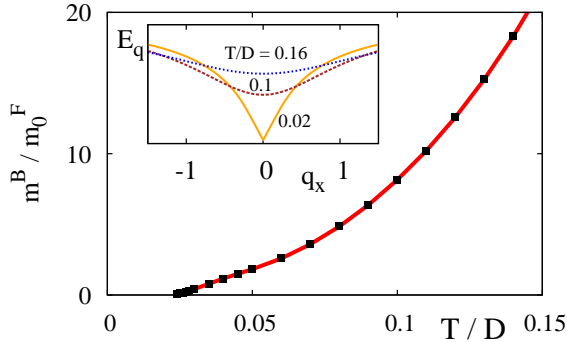

FIG. 2: (Color online) Enhancement of the preformed pairs' mobility with the decreasing temperature obtained from self-consistent solution of the flow equations (7,8) for the constant charge concentration $n_{tot} = 2$. Inset shows the bosonic dispersion $\tilde{E}_\mathbf{q}$ for a few representative temperatures.

The high $T_c$ cuprate oxides are nearly two-dimensional materials where the superconductivity is driven in $CuO_2$ planes. For this reason we can interpret the temperature dependent mass $m^B$ of the preformed pairs as a quantity related with the residual Meissner effect in the reduced dimensions [46]. This aspect has been recently emphasized by the ETH group [47] within the Quantum Monte Carlo studies of the present model (1). Following the same routine we show in figure 3 the diamagnetic magnetization $M_d(T)$ obtained from the continuous unitary transformation for the 2-dimensional case with $n_{tot} = 2$. We can notice that the increasing mobility of the preformed pairs substantially enhances the magnetization. This behavior can be independently explained by the direct calculation of the current-current response function (discussed in the next section).

## IV. EFFECT OF THE PREFORMED PAIRS ON THE RESPONSE FUNCTION

The residual Meissner effect and the conductivity can be obtained from the response function $\Pi_{\alpha,\beta}(\mathbf{q},\tau) \equiv$ $-\langle \hat{T}_\tau \hat{j}_{\mathbf{q},\alpha}(\tau) \hat{j}_{-\mathbf{q},\beta} \rangle$ [where $\alpha,\beta$ denote the Cartesian $x,y,z$ coordinates] with the current operator defined as

$$\hat{\mathbf{j}}_\mathbf{q} = \sum_\mathbf{k} \mathbf{v}_{\mathbf{k}+\frac{\mathbf{q}}{2}} \sum_{\sigma=\uparrow,\downarrow} \hat{c}^\dagger_{\mathbf{k},\sigma} \hat{c}_{\mathbf{k}+\mathbf{q},\sigma} \qquad (9)$$

and velocity $\mathbf{v}_\mathbf{k} = \hbar^{-1}\nabla_\mathbf{k}\varepsilon_\mathbf{k}$. Within the continuous unitary transformation it is convenient to compute the current-current response function $\Pi_{\alpha,\beta}(\mathbf{q},i\nu)$ using the statistical averages with respect to the diagonalized Ha-miltonian $\hat{H}(\infty)$. This, however, requires that the current operator (9) has to be analyzed in the same transformation routine as the Hamiltonian. Some technical details concerning derivation of $\Pi_{\alpha,\beta}(\mathbf{q},i\nu)$ are outlined in the appendix. In the asymptotic limit $l \rightarrow \infty$ we obtain two contributions to the response function, from: (a) the BE condensed pairs and (b) the finite momentum preformed pairs (see Fig. 1 of Ref. [28]). Explicit form of the response function is given by [28]

$$\begin{aligned}\Pi_{\alpha,\beta}(\mathbf{q},i\nu) &= \sum_\mathbf{k} v_{\mathbf{k}+\frac{\mathbf{q}}{2},\alpha} v_{\mathbf{k}+\frac{\mathbf{q}}{2},\beta} \left\{ \mathcal{N}_{\mathbf{k},\mathbf{q}} \left[ f_{FD}(\tilde{\xi}_{\mathbf{k}+\mathbf{q}}) - f_{FD}(\tilde{\xi}_\mathbf{k}) \right] \left[ \frac{1}{i\nu + \tilde{\xi}_{\mathbf{k}+\mathbf{q}} - \tilde{\xi}_\mathbf{k}} - \frac{1}{i\nu - \tilde{\xi}_{\mathbf{k}+\mathbf{q}} + \tilde{\xi}_\mathbf{k}} \right] \right. \\ &+ \frac{1}{\sqrt{N}} \sum_{\mathbf{k}'} \mathcal{M}_{\mathbf{k},\mathbf{k}',\mathbf{q}} \left( \left[ f_{BE}(\tilde{E}_{\mathbf{k}-\mathbf{k}'}) - f_{BE}(\tilde{\xi}_{\mathbf{k}+\mathbf{q}} + \tilde{\xi}_{\mathbf{k}'}) \right] \frac{1 - f_{FD}(\tilde{\xi}_{\mathbf{k}+\mathbf{q}}) - f_{FD}(\tilde{\xi}_{\mathbf{k}'})}{i\nu - (\tilde{\xi}_{\mathbf{k}+\mathbf{q}} + \tilde{\xi}_{\mathbf{k}'} - \tilde{E}_{\mathbf{k}-\mathbf{k}'})} \right.\end{aligned}$$

$$- \left[ f_{BE}(\tilde{E}_{\mathbf{k}-\mathbf{k}'}) - f_{BE}(\tilde{\xi}_{\mathbf{k}'+\mathbf{q}} + \tilde{\xi}_{\mathbf{k}}) \right] \frac{1 - f_{FD}(\tilde{\xi}_{\mathbf{k}'+\mathbf{q}}) - f_{FD}(\tilde{\xi}_{\mathbf{k}})}{i\nu + (\tilde{\xi}_{\mathbf{k}'+\mathbf{q}} + \tilde{\xi}_{\mathbf{k}} - \tilde{E}_{\mathbf{k}-\mathbf{k}'})} \bigg) \bigg\}, \quad (10)$$

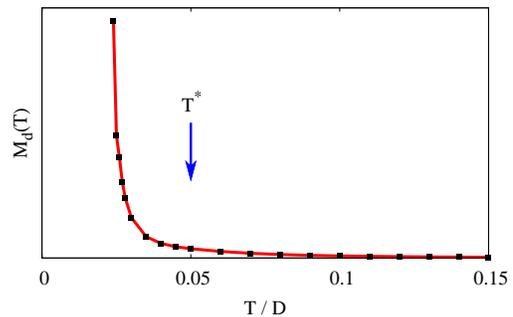

FIG. 3: (Color online) Residual diamagnetism induced above $T_c$ by the preformed electron pairs. Magnetization has been computed using $M_d(T) \propto 1/m^B(T)$ suitable for the 2-dimensional hard-core boson gas [46] in analogy to QMC studies [47] of the present model.

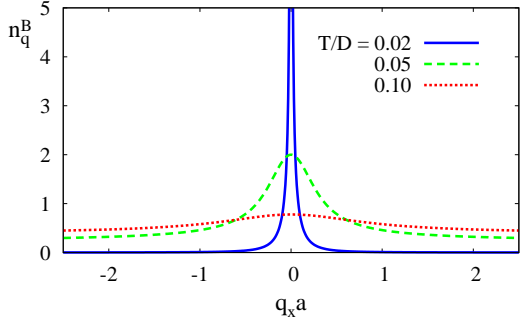

FIG. 4: (Color online) Gradual accumulation of the preformed pairs at low-momenta, leading to appearance of the Drude peak in a.c. conductivity.

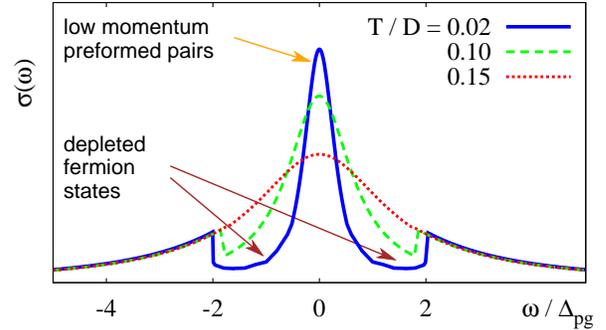

FIG. 5: (Color online) The dynamic conductivity $\sigma(\omega)$ revealing the Drude peak caused by the low momentum preformed pairs and the optical gap $|\omega| \leq \Delta_{pg}$ due to the depleted single particle states (i.e. pseudogap). Energy $\omega$ is expressed in units of the pseudogap $\Delta_{pg}$ at low temperature $T = 0.02D$.

with the Fermi-Dirac $f_{FD}(\omega) = [\exp(\omega/k_B T) + 1]^{-1}$ and Bose-Einstein $f_{BE}(\omega) = [\exp(\omega/k_B T) - 1]^{-1}$ functions, respectively. The coefficients

$$\begin{aligned}
\mathcal{N}_{\mathbf{k},\mathbf{q}} &\equiv \tilde{\mathcal{A}}_{\mathbf{k},\mathbf{q}} \tilde{\mathcal{A}}_{\mathbf{k}+\mathbf{q},-\mathbf{q}} + \tilde{\mathcal{A}}_{\mathbf{k},\mathbf{q}} \tilde{\mathcal{B}}_{-\mathbf{k},-\mathbf{q}} \\
&+ \tilde{\mathcal{A}}_{-\mathbf{k},-\mathbf{q}} \tilde{\mathcal{B}}_{\mathbf{k},\mathbf{q}} + \tilde{\mathcal{B}}_{\mathbf{k},\mathbf{q}} \tilde{\mathcal{B}}_{\mathbf{k}+\mathbf{q},-\mathbf{q}} \quad (11) \\
\mathcal{M}_{\mathbf{k},\mathbf{k}',\mathbf{q}} &\equiv \left( \tilde{\mathcal{D}}_{\mathbf{k}+\mathbf{q},-(\mathbf{k}'+\mathbf{q}),-\mathbf{q}} - \tilde{\mathcal{F}}_{\mathbf{k}+\mathbf{q},-(\mathbf{k}'+\mathbf{q}),-\mathbf{q}} \right) \\
&\times \left( \tilde{\mathcal{D}}_{\mathbf{k},-\mathbf{k}',\mathbf{q}} - \tilde{\mathcal{F}}_{\mathbf{k},-\mathbf{k}',\mathbf{q}} \right) \quad (12)
\end{aligned}$$

denote the asymptotic values of parameters introduced in the $l$-dependent current operator (A1). This response function (10) generalizes the standard BCS result [29, 45] taking into account the finite momentum preformed pairs $n^B_{\mathbf{q}\neq\mathbf{0}}$. They enter the response function through the terms proportional to $\mathcal{M}_{\mathbf{k},\mathbf{k}',\mathbf{q}}$ and their influence leads to appearance of the Drude peak in the subgap optical conductivity (Fig. 4).

Let us remark, that in the superconducting state the electrodynamic response is dominated by the BE-condensed ($\mathbf{q}=\mathbf{0}$) pairs. In such situation the coefficients (11,12) simplify to the usual BCS coherence factors $\mathcal{N}_{\mathbf{k},\mathbf{q}} = (u_{\mathbf{k}+\mathbf{q}} u_{\mathbf{k}} + v_{\mathbf{k}+\mathbf{q}} v_{\mathbf{k}})^2$ and $\mathcal{M}_{\mathbf{k},\mathbf{k},\mathbf{q}} \frac{1}{\sqrt{N}} f_{BE}(\tilde{E}_{\mathbf{q}=\mathbf{0}}) = (u_{\mathbf{k}+\mathbf{q}} v_{\mathbf{k}} - v_{\mathbf{k}+\mathbf{q}} u_{\mathbf{k}})^2$ [28]. Since the preformed pairs are concentrated in the low momentum (long-wavelength) states (see Fig. 4), therefore some of these BCS features can be preserved also in the pseudogap state above $T_c$.

## V. FLUCTUATION CONDUCTIVITY ABOVE $T_c$

We now analyze how the preformed pairs show up in the a.c. (dynamic) conductivity defined by [45]

$$\sigma_{\alpha,\beta}(\mathbf{q},\omega) = -\frac{1}{\omega} \, \text{Im} \left[ \Pi_{\alpha,\beta}(\mathbf{q},\omega) \right]. \quad (13)$$

For specific considerations we focus on 2-dimensional lattice version of the boson-fermion model (1), characterized by the tight-binding dispersion $\xi_{\mathbf{k}} = -2t \left[ \cos(k_x a) + \cos(k_y a) \right] - \mu$. In this expression t is the hopping integral, and the bandwidth $D \equiv 8t$ is used as a unit for the energies. We assume that (initially) the preformed pairs are dispersionless (localized) $E_{\mathbf{q}} = \Delta_B - 2\mu$ but they acquire some itinerancy due to boson-fermion coupling $g_{\mathbf{k},\mathbf{p}}$. We have constructed the numerical codes using the following set of parameters $\Delta_B = 0$, $g_{\mathbf{k},\mathbf{k}'} = 0.8D$. We have determined the chemical potential $\mu(T)$ keeping the fixed charge concentration $n_{tot} = 2$.

We solved the differential equations (6-8) along with the flow equations (A2,A5) for the parametrized current operator (A1). We have covered the Brillouin zone by a mesh of 500×500 equidistant points and solved the coupled differential equations using the Runge-Kutta algorithm. The flow parameter $l \rightarrow l + \delta l$ has been changed with the flexible increment $\delta l$ adjusted in order to control the ongoing renormalizations. In the initial stage of transformation we used $\delta l = 0.0001 D^{-2}$, and later on we increased its value as discussed by us in Refs [37, 42].

To avoid summations of the sharp delta functions we have imposed a small imaginary part in the analytical continuation $\Pi_{\alpha,\beta}(\mathbf{q}, i\nu) \rightarrow \Pi_{\alpha,\beta}(\mathbf{q}, \omega + i\tau^{-1})$. Roughly speaking $\tau$ can be regarded as some phenomenological scattering time, which we assume to be constant for the discussed temperature regime. In the normal state the dynamic conductivity is characterized by the Drude model behavior $\sigma(\omega) = \sigma_N / \left(1 + \omega^2 \tau^2\right)$ with the d.c. conductivity $\sigma_N = ne^2 \tau / m^F$. It obeys the important f-sum rule $\int_{-\infty}^{\infty} d\omega \sigma(\omega) = \pi\, ne^2 / m^F$.

Figure 5 shows the a.c. conductivity obtained for the dirty limit $\sigma(\omega) \equiv \frac{1}{N} \sum_{\mathbf{q}} \sigma_{x,x}(\mathbf{q},\omega)$. Upon lowering temperature we observe that: (i) depletion of the single particle (fermion) states near the Fermi level induces [through the terms (11)] the optical gap over energy regime $\omega \in \langle -2\Delta_{pg}; 2\Delta_{pg}\rangle$, (ii) accumulation of the low-momentum preformed pairs (bosons) contributes [via the terms (12)] more and more spectral weight to the Drude peak. Transfer of this spectral weight goes hand in hand with a gradual emergence of the diamagnetism (Fig. 3) in very much the same way as it does in the symmetry-broken superconducting state [29].

The ongoing transfer of the optical weight has the indirect consequence on temperature variation of the d.c. conductivity $\sigma(0)$. We observe that d.c. conductivity is substantially enhanced with decreasing temperature in the pseudogap regime. This "fluctuation enhanced conductivity" is well known experimentally. As an example we show in Fig. 6 the temperature dependent resistivity $\rho = 1/\sigma(0)$ of the bismuth cuprate superconductors. Subtracting the normal state value $\rho_n$ we can notice that the reduced resistivity (enhanced conductivity) starts well above the transition temperature, already at $T^* \simeq 2.2 T_c$. As concerns the optical gap of the a.c. conductivity this effect has been reported for various families of the cuprate superconductors [16] in the temperature and doping regime corresponding to the residual Meissner effect [17]. Similar fluctuation effects have been observed also in the strongly disordered thin classical superconductors [18].

## VI. SUMMARY

We have studied influence of the preformed local pairs on the diamagnetic response and the conductivity in the pseudogap region above $T_c$. For specific considerations we have used the boson-fermion model, describing the itinerant electrons interacting via the Andreev-type scattering with the preformed local pairs. We have shown that a gradual suppression of the single particle states near the Fermi energy is accompanied by an increasing mobility of the preformed pairs (Fig. 2). The latter effect leads in turn to some fragile diamagnetic response of the system (Fig. 3). We have further supported this result from analysis of the preformed pairs contribution to the current-current response function, that has been determined within the flow equation procedure beyond the perturbative scheme.

We have also investigated the dynamic conductivity and found that the suppressed fermionic spectrum induces the optical gap in the infrared regime $|\omega| \leq 2\Delta_{pg}$ while the accumulation of the low-momentum preformed pairs gives rise to the Drude-like peak. Upon lowering the temperature there is more and more spectral weight transferred to the Drude peak at expense of deepening the optical gap. This processes driven by the low momentum preformed pairs does amplify (via f-sum rule) the d.c. conductivity. Finally, we have confronted such fluctuation conductivity with the experimental data obtained for the Bi2223 cuprate superconductors.

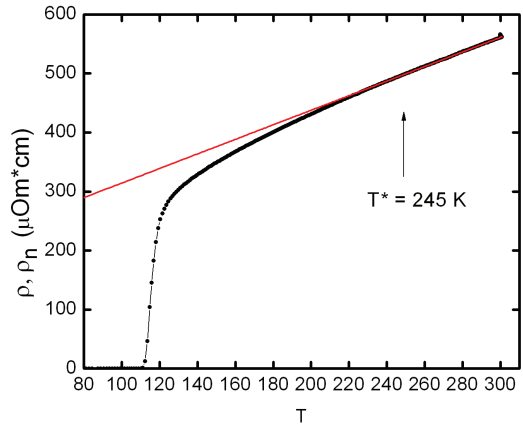

FIG. 6: (Color online) Temperature dependence of the d.c. resistivity of Bi2223 cuprate superconductors. Notice that the fluctuation conductivity occurs below $T^* \approx 2.2 T_c$.

### Acknowledgments

T.D. acknowledges discussions with J. Ranninger and R. Micnas. This work is supported by the National Science Centre in Poland through the project DEC-2014/13/B/ST3/04451 (TD).

## APPENDIX A: TRANSFORMATION OF THE CURRENT OPERATOR

We briefly outline here the continuous transformation for the current operator (9). We individually study both spin contributions $\hat{\mathbf{j}}_{\mathbf{q}}^{\sigma} = \sum_{\mathbf{k}} \mathbf{v}_{\mathbf{k}+\frac{\mathbf{q}}{2}} \hat{c}_{\mathbf{k},\sigma}^{\dagger} \hat{c}_{\mathbf{k}+\mathbf{q},\sigma}$ because their evolution with respect to $l$ is slightly different. From the initial ($l=0$) derivative $\frac{d}{dl}\hat{\mathbf{j}}_{\mathbf{q}}^{\sigma}(l) = [\hat{\eta}(l), \hat{\mathbf{j}}_{\mathbf{q}}^{\sigma}(l)]$ we conclude the following ($l$-dependent) parametrization

$$\hat{\mathbf{j}}_{\mathbf{q}}^{\uparrow}(l) = \sum_{\mathbf{k}} \mathbf{v}_{\mathbf{k}+\frac{\mathbf{q}}{2}} \Big( \mathcal{A}_{\mathbf{k},\mathbf{q}}(l) \hat{c}_{\mathbf{k},\uparrow}^{\dagger} \hat{c}_{\mathbf{k}+\mathbf{q},\uparrow}$$
$$+ \mathcal{B}_{\mathbf{k},\mathbf{q}}(l) \hat{c}_{-\mathbf{k},\downarrow} \hat{c}_{-(\mathbf{k}+\mathbf{q}),\downarrow}^{\dagger} + \sum_{\mathbf{p}} \mathcal{D}_{\mathbf{k},\mathbf{p},\mathbf{q}}(l) \hat{b}_{\mathbf{k}+\mathbf{p}} \hat{c}_{\mathbf{k},\uparrow}^{\dagger} \hat{c}_{\mathbf{p}-\mathbf{q},\downarrow}^{\dagger}$$

$$+ \sum_{\mathbf{p}} \mathcal{F}_{\mathbf{k},\mathbf{p},\mathbf{q}}(l) \hat{b}^{\dagger}_{\mathbf{k}+\mathbf{p}} \hat{c}_{\mathbf{p},\downarrow} \hat{c}_{\mathbf{k}+\mathbf{q},\uparrow} \Bigg). \tag{A1}$$

The other spin contribution $\hat{\mathbf{j}}^{\downarrow}_{\mathbf{q}}(l)$ has the coefficient $\mathcal{D}_{\mathbf{k},\mathbf{p},\mathbf{q}}(l)$ interchanged with $-\mathcal{F}_{\mathbf{k},\mathbf{p},\mathbf{q}}(l)$ and *vice versa*. The new parameters are subject to the boundary conditions $\mathcal{A}_{\mathbf{k},\mathbf{q}}(0) = 1$ and $\mathcal{B}_{\mathbf{k},\mathbf{q}}(0) = \mathcal{D}_{\mathbf{k},\mathbf{p},\mathbf{q}}(0) = \mathcal{F}_{\mathbf{k},\mathbf{p},\mathbf{q}}(0) = 0$. Let us remark here, that restricting only to the BE condensed pairs $\hat{b}^{(\dagger)}_{\mathbf{k}+\mathbf{p}} = \hat{b}^{(\dagger)}_{\mathbf{0}} \delta_{\mathbf{p},-\mathbf{k}}$ the constraint (A1) exactly reproduces the standard BCS solution [28]. For arbitrary case we can derive from the operator equation $\frac{d\hat{\mathbf{j}}^{\sigma}_{\mathbf{q}}(l)}{dl} = [\hat{\eta}(l), \hat{\mathbf{j}}^{\sigma}_{\mathbf{q}}(l)]$ the following set of flow equations

$$\frac{d\mathcal{A}_{\mathbf{k},\mathbf{q}}(l)}{dl} = \sum_{\mathbf{p}} \left[ \alpha_{\mathbf{k}+\mathbf{q},\mathbf{p}-\mathbf{q}}(l) \mathcal{D}_{\mathbf{k},\mathbf{p},\mathbf{q}}(l) \left( n^F_{\mathbf{p}-\mathbf{q},\sigma} + n^B_{\mathbf{k}+\mathbf{p}} \right) \right.$$
$$\left. + \alpha_{\mathbf{k},\mathbf{p}}(l) \mathcal{F}_{\mathbf{k},\mathbf{p},\mathbf{q}}(l) \left( n^F_{\mathbf{p},\sigma} + n^B_{\mathbf{k}+\mathbf{p}} \right) \right] \tag{A2}$$

$$\frac{d\mathcal{B}_{\mathbf{k},\mathbf{q}}(l)}{dl} = -\sum_{\mathbf{p}} \left[ \alpha_{\mathbf{k},\mathbf{p}}(l) \mathcal{D}_{-\mathbf{p},-\mathbf{k},\mathbf{q}}(l) \left( n^F_{\mathbf{p},\sigma} + n^B_{\mathbf{k}+\mathbf{p}} \right) \right.$$
$$\left. + \alpha_{\mathbf{k}+\mathbf{q},\mathbf{p}-\mathbf{q}}(l) \mathcal{F}_{-\mathbf{p},-\mathbf{k},\mathbf{q}}(l) \left( n^F_{\mathbf{p}-\mathbf{q},\sigma} + n^B_{\mathbf{k}+\mathbf{p}} \right) \right] \tag{A3}$$

$$\frac{d\mathcal{D}_{\mathbf{k},\mathbf{p},\mathbf{q}}(l)}{dl} = -\alpha_{\mathbf{k}+\mathbf{q},\mathbf{p}-\mathbf{q}}(l) \mathcal{A}_{\mathbf{k},\mathbf{q}}(l) + \alpha_{\mathbf{k},\mathbf{p}}(l) \mathcal{B}_{-\mathbf{p},\mathbf{q}}(l), \tag{A4}$$

$$\frac{d\mathcal{F}_{\mathbf{k},\mathbf{p},\mathbf{q}}(l)}{dl} = -\alpha_{\mathbf{k},\mathbf{p}}(l) \mathcal{A}_{\mathbf{k},\mathbf{q}}(l) + \alpha_{\mathbf{k}+\mathbf{q},\mathbf{p}-\mathbf{q}}(l) \mathcal{B}_{-\mathbf{p},\mathbf{q}}(l). \tag{A5}$$

These complex equations can be either solved numerically or (with some compromise) analytically. The lowest order estimation of the coefficients $\mathcal{A} - \mathcal{F}$ is feasible for instance if we neglect renormalizations of the fermion and boson energies on the right hand side of the equations (A2-A5). Substituting the exponential scaling $g_{\mathbf{k},\mathbf{p}}(l) \simeq g_{\mathbf{k},\mathbf{p}} e^{-(\xi_{\mathbf{k}}+\xi_{\mathbf{p}}-E_{\mathbf{k}+\mathbf{p}})^2 l}$ these flow equations (A2-A5) can be solved iteratively, starting from the initial (boundary) conditions. Thus estimated coefficients (A4,A5) are given by [28]

$$\tilde{\mathcal{A}}_{\mathbf{k},\mathbf{q}} \simeq 1 - \frac{1}{2} \sum_{\mathbf{p}} \left[ \frac{\left( n^F_{\mathbf{p}} + n^B_{\mathbf{k}+\mathbf{p}} \right) |g_{\mathbf{k},\mathbf{p}}|^2}{(\xi_{\mathbf{k}} + \xi_{\mathbf{p}} - E_{\mathbf{k}+\mathbf{p}})^2} \right.$$
$$\left. + \frac{\left( n^F_{\mathbf{p}-\mathbf{q}} + n^B_{\mathbf{k}+\mathbf{p}} \right) |g_{\mathbf{k}+\mathbf{q},\mathbf{p}-\mathbf{q}}|^2}{(\xi_{\mathbf{k}+\mathbf{q}} + \xi_{\mathbf{p}-\mathbf{q}} - E_{\mathbf{k}+\mathbf{p}})^2} \right] \tag{A6}$$

$$\tilde{\mathcal{D}}_{\mathbf{k},\mathbf{p},\mathbf{q}} \simeq -\frac{g_{\mathbf{k}+\mathbf{q},\mathbf{p}-\mathbf{q}}}{\xi_{\mathbf{k}+\mathbf{q}} + \xi_{\mathbf{p}-\mathbf{q}} - E_{\mathbf{k}+\mathbf{p}}}, \tag{A7}$$

$$\tilde{\mathcal{F}}_{\mathbf{k},\mathbf{p},\mathbf{q}} \simeq -\frac{g_{\mathbf{k},\mathbf{p}}}{\xi_{\mathbf{k}} + \xi_{\mathbf{p}} - E_{\mathbf{k}+\mathbf{p}}}. \tag{A8}$$

and

$$\tilde{\mathcal{B}}_{\mathbf{k},\mathbf{q}} \simeq \sum_{\mathbf{p}} g_{\mathbf{k},\mathbf{p}} g_{\mathbf{k}+\mathbf{q},\mathbf{p}-\mathbf{q}} \times \tag{A9}$$
$$\left[ \frac{n^F_{\mathbf{p}} + n^B_{\mathbf{k}+\mathbf{p}}}{X_{\mathbf{k}+\mathbf{q},\mathbf{p}-\mathbf{q}}} \left( \frac{1}{X_{\mathbf{k},\mathbf{p}}} - \frac{X_{\mathbf{k},\mathbf{p}}}{X^2_{\mathbf{k},\mathbf{p}} + X^2_{\mathbf{k}+\mathbf{q},\mathbf{p}-\mathbf{q}}} \right) + \right.$$
$$\left. \frac{n^F_{\mathbf{p}-\mathbf{q}} + n^B_{\mathbf{k}+\mathbf{p}}}{X_{\mathbf{k},\mathbf{p}}} \left( \frac{1}{X_{\mathbf{k}+\mathbf{q},\mathbf{p}-\mathbf{q}}} - \frac{X_{\mathbf{k}+\mathbf{q},\mathbf{p}-\mathbf{q}}}{X^2_{\mathbf{k},\mathbf{p}} + X^2_{\mathbf{k}+\mathbf{q},\mathbf{p}-\mathbf{q}}} \right) \right]$$

where $X_{\mathbf{k},\mathbf{p}} \equiv \xi_{\mathbf{k}} + \xi_{\mathbf{p}} - E_{\mathbf{k}+\mathbf{p}}$.